\documentclass[a4paper]{jpconf}
\usepackage{graphicx}
\usepackage{amssymb}
\usepackage{amsmath}
\usepackage{siunitx}

\begin{document}
\title{Determination of the Charge per Micro-Bunch of a Self-Modulated Proton Bunch using a Streak Camera}

\author{A.-M. Bachmann$^{1,2,3}$, P. Muggli$^{2,1}$}
\address{$^1$ CERN, Geneva, Switzerland}
\address{$^2$ Max-Planck Institute for Physics, Munich, Germany}
\address{$^3$ Technical University Munich, Munich, Germany}
\ead{anna-maria.bachmann@cern.ch}
%\address{Production Editor, \jpcs, \iopp, Dirac House, Temple Back, Bristol BS1~6BE, UK}

%\ead{jacky.mucklow@iop.org}

\begin{abstract}
    The Advanced Wakefield Experiment (AWAKE) develops the first plasma wakefield accelerator with a high-energy proton bunch as driver. The \SI{400}{\giga\electronvolt} bunch from CERN Super Proton Synchrotron (SPS) propagates through a \SI{10}{\m} long rubidium plasma, ionized by a \SI{4}{\tera\watt} laser pulse co-propagating with the proton bunch. The relativistic ionization front seeds a self-modulation process. The seeded self-modulation transforms the bunch into a train of micro-bunches resonantly driving wakefields. We measure the density modulation of the bunch, in time, with a streak camera with picosecond resolution. The observed effect corresponds to alternating focusing and defocusing fields. We present a procedure recovering the charge of the bunch from the experimental streak camera images containing the charge density. These studies are important to determine the charge per micro-bunch along the modulated proton bunch and to understand the wakefields driven by the modulated bunch. 
\end{abstract}

\section{Introduction}
\label{sec:intro}
\begin{figure}[h!]
    \includegraphics[width=25pc,trim={0cm 25.9cm 8cm 0cm},clip]{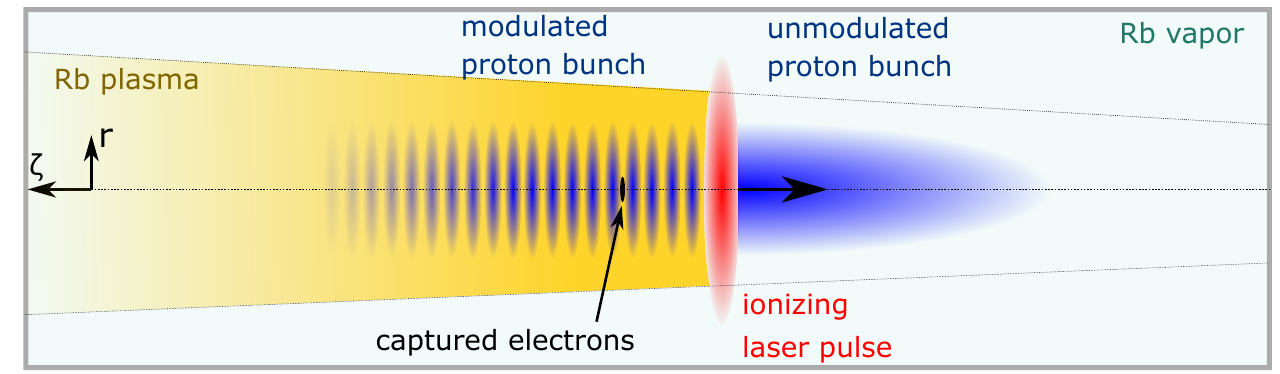}
    \hspace{2pc}%
    \begin{minipage}[b]{9pc}
        \caption{\label{label} Sketch of the AWAKE principle, using a proton bunch as plasma wakefield driver to accelerate externally injected electrons}
        \label{fig:sketchofawakeprinciple}
    \end{minipage}
\end{figure}  
    AWAKE uses the CERN SPS proton bunch as a plasma wakefield driver. The bunch propagates through \SI{10}{\m} of plasma, created by laser ionization of a rubidium (Rb) vapor. The laser pulse co-propagates with the proton bunch, seeding the self-modulation with the relativistic ionization front, i.e. with the abrupt beam plasma interaction within the bunch \cite{AWAKE}. Along the plasma (with a density of $n_{pe}=2\cdot 10^{14} \, \textnormal{cm}^{-3}$ for the measurements reported here) the long proton bunch ($\sigma_{\zeta} \approx 9 \, \textnormal{cm}$) divides into micro-bunches, spaced by the plasma wavelength ($\lambda_{pe} \approx 2.4 \, \textnormal{mm}$) \cite{Marlene,Karl}. The micro-bunches resonantly drive wakefields in the plasma. The wakefields accelerate an injected electron witness bunch \cite{Nature}. The principle of the experiment is sketched in Figure \ref{fig:sketchofawakeprinciple}.\\
    In the following we present a method to determine the charge in each micro-bunch from time-resolved images of the proton bunch transverse distribution. The images are produced by a streak camera.
    
\section{Method}
\label{sec:method} 
In this section we explain the analysis of the streak camera images \cite{KarlSc} applied for the determination of the charge per micro-bunch. 
\subsection{Streak Camera as Diagnostic}
    \begin{figure}[h!]
        \begin{minipage}{24pc}
            \includegraphics[width=24pc,trim={0cm 23.3cm 0cm 0cm},clip]{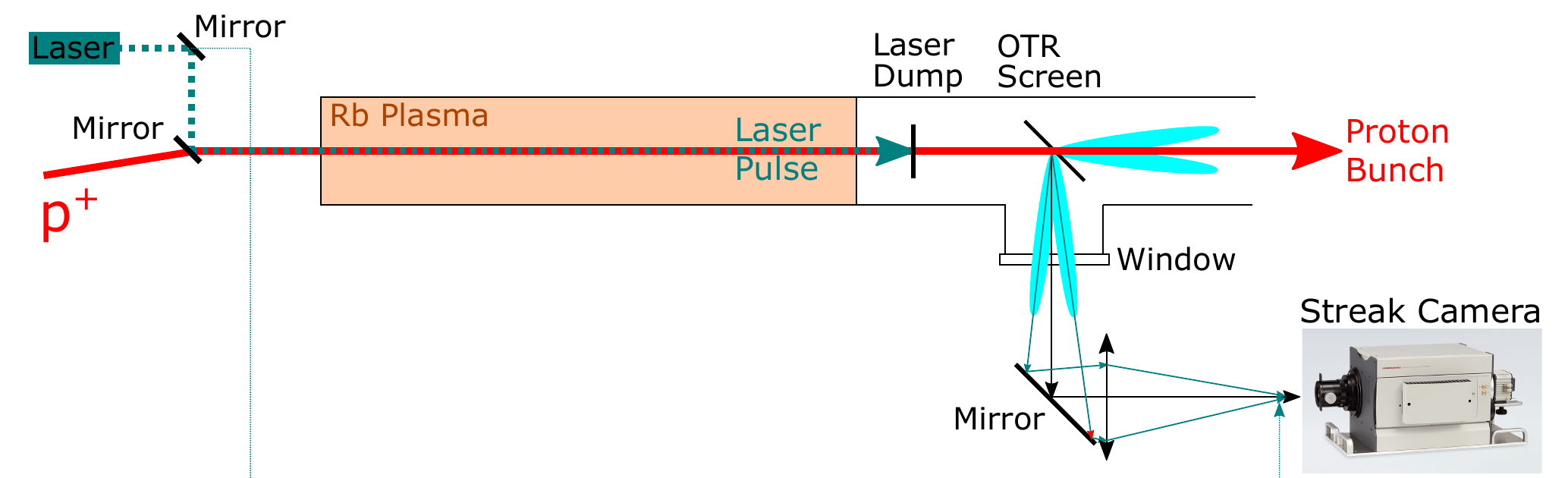}
            \caption{\label{label} Transport of OTR light of a modulated proton bunch to streak camera at AWAKE (not to scale)}
            \label{fig:transportOTR}
        \end{minipage}\hspace{2pc}%
        \begin{minipage}{11pc}
            \includegraphics[width=13pc,trim={0cm 20cm 0cm 0cm},clip]{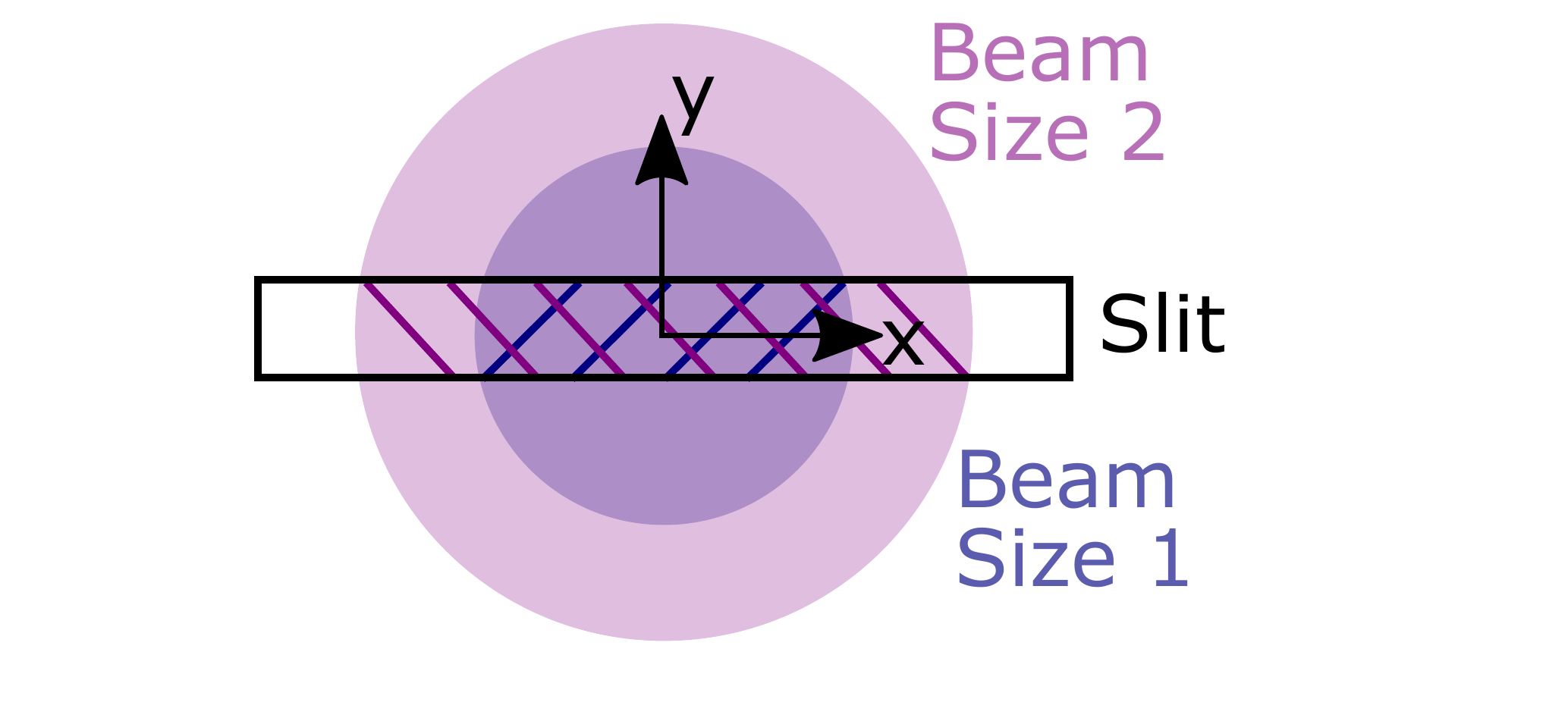}
            \caption{\label{label} Light collection of a streak camera for different signal widths}
            \label{fig:sketchslitsc}
        \end{minipage} 
    \end{figure}
    After the Rb plasma the modulated proton bunch propagates through an optical transition radiation (OTR) screen (\SI{280}{\micro\m} thick Silicon wafer coated with \SI{1}{\micro\m} thick mirror-finished aluminium), placed \SI{3.5}{\m} after the plasma exit \cite{Karl}, see Figure \ref{fig:transportOTR}. We collect the backwards emitted OTR that contains the spatio-temporal information of the proton bunch charge distribution and transport it to a streak camera (Hamamatsu C10910-05 model, 16-bit, 2048 x 2048 pixel ORCA-Flash4.0 CMOS sensor, binned to 512 x 512 pixels for streak operation). The imaging system has a limited aperture ($\pm 4 \, \textnormal{mm}$ in Figure \ref{fig:example_streakcameraimage_laseroff} and \ref{fig:example_streakcameraimage_laseron} and later). We operate the camera with a slit width of \SI{20}{\micro\m}, an MCP gain of $40$ and a time window of $73 \, \textnormal{ps}$. The time resolution is $\approx 1 \, \textnormal{ps}$ in this time window. Light is collected by the streak camera through a slit for temporal resolution. Thus, for a cylindrically symmetric light signal, as the transverse image of the proton bunch, with a size larger than the slit width, the larger the size, the smaller the fraction of light that is transmitted through the slit (Figure \ref{fig:sketchslitsc}). The streak camera image thus contains information about the bunch charge density and not the charge. 
    
\subsection{Streak Camera Images}
    \begin{figure}[h!]
        \begin{minipage}{18pc}
            \vspace{-0.5cm}
            \includegraphics[width=18pc,trim={0.9cm 0cm 1.5cm 0cm},clip]{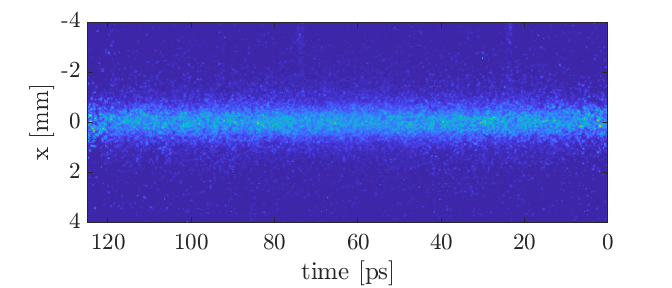}
            \caption{\label{label} Stitched streak camera image of the proton bunch after propagation without plasma (bunch front at $t=0$). The marker laser pulses used for stitching are at the top edge of the image. The bunch density is almost uniform along the bunch.}
            \label{fig:example_streakcameraimage_laseroff}
        \end{minipage}\hspace{2pc}%
        \begin{minipage}{18pc}
            \includegraphics[width=18pc,trim={0.9cm 0cm 1.5cm 0cm},clip]{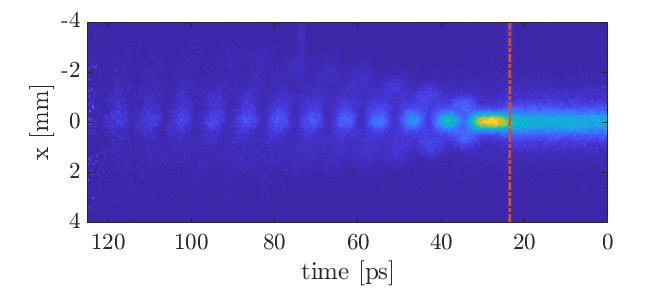}
            \caption{\label{label} Stitched streak camera image of the proton bunch with plasma. The red line shows the location of the ionizing laser pulse ($t=24 \, \textnormal{ps}$). The head of the bunch ($t<24 \, \textnormal{ps}$) propagated through vapor. The tail ($t>24 \, \textnormal{ps}$) propagated through plasma and is self-modulated.}
            \label{fig:example_streakcameraimage_laseron}
        \end{minipage} 
    \end{figure}
    
%We interpolate the original streak camera image to linearize the time axis and subtract a background of the camera (dark noise). We average images of 20 events per time window for the modulated bunch that propagated through plasma, 2 for the unmodulated bunch when propagating through vapor, when the ionizing laser pulse was blocked. We adjust for the jitter of the streak camera trigger by sending the bleed through of the ionizing laser pulse of a mirror in the transport line as a timing reference onto the streak camera (see turquoise line in sketch \ref{fig:transportOTR}), missteered to cover only the left half of the image and minimize the overlap with the proton bunch signal (see e.g. fig. \ref{fig:example_streakcameraimage_laseroff}). Due to the high intensity of the pulse, thus space charge effects, and time resolution of the streak camera, the signal is broadened from approx $120 \, \textnormal{fs}$ (laser pulse duration) to $ps$-scale. We append a second time window shifting the time reference pulse with a delay stage and the timing of the streak camera trigger. We average the images normalizing each image with the incoming total proton bunch population of the event, measured by a beam current transformer inside the SPS ring. 

The streak camera produces a time resolved image of the proton bunch transverse charge distribution \cite{KarlSc}. The temporal evolution of the streak voltage leads to a time interval per pixel that varies along the image. Therefore we interpolate the original image to linearize the time axis. We subtract from each image a background image, obtained by averaging seven images without proton bunch. Images are weighted by the measured incoming proton bunch population. We acquire two series of images (each 20 images with plasma, two images without plasma) with $\approx 50 \, \textnormal{ps}$ delay between series. Together with the proton bunch OTR, we send to the streak camera a replica of the ionizing laser pulse ($\approx 120 \, \textnormal{fs}$ long) that we also delay by $50 \, \textnormal{ps}$ for each series. This reference laser pulse is sent onto the edge of the image to minimize the overlap with the proton bunch signal (see top edge on Figure \ref{fig:example_streakcameraimage_laseroff} and \ref{fig:example_streakcameraimage_laseron}). With this laser pulse time reference we can sum images in a series with the same time delay, despite the $\approx 20 \, \textnormal{ps}$ trigger jitter of the streak camera. We stitch the series together to obtain long time scale images with short time scale resolution. This method is described in reference \cite{FBMarker} of these proceedings.\\
The result without plasma is shown in Figure \ref{fig:example_streakcameraimage_laseroff} and with plasma in Figure \ref{fig:example_streakcameraimage_laseron}. Here $t=0$ corresponds to $6 \, \textnormal{ps}$ behind the proton bunch center, with a bunch length of $\sigma_{\zeta} = 300 \, \textnormal{ps}$ and a total population of $N_{p^+} = (2.8 \pm 0.2) \cdot 10^{11}$. The transverse center of the bunch $x=0$ was determined as the peak of a Gaussian fit of the unmodulated head of the bunch, before the ionizing laser pulse ($t < 24 \, \textnormal{ps}$). After propagation through Rb vapor, the bunch charge distribution is uniform (Figure \ref{fig:example_streakcameraimage_laseroff}). After propagating through plasma (Figure \ref{fig:example_streakcameraimage_laseron}), the proton bunch is self-modulated. One can see the micro-bunches along the propagation axis as well as defocused protons in between. The image shows that the charge density of the micro-bunches decreases along the bunch. In the following we present a method to determine the charge per micro-bunch for a change in width (radius) along the bunch.
    
\subsection{Micro-Bunch Temporal Structure}

    \begin{figure}[h!]
        \begin{minipage}{18pc}
            \includegraphics[width=18pc,trim={0.5cm 0cm 0cm 0cm},clip]{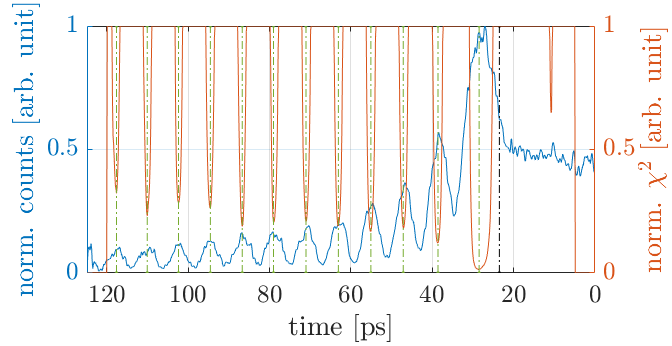}
            \caption{\label{label} Determination of the longitudinal center of micro-bunches with the central projection of the modulated bunch (blue solid line), the ionizing laser pulse timing (black dashed line), the weighted distance squared function $\chi^2$ (orange solid line) and its minima (green dashed line)}
            \label{fig:findMBsc}
        \end{minipage}\hspace{2pc}%
        \begin{minipage}{18pc}
            \includegraphics[width=18pc,trim={0.5cm 0cm 0cm 0cm},clip]{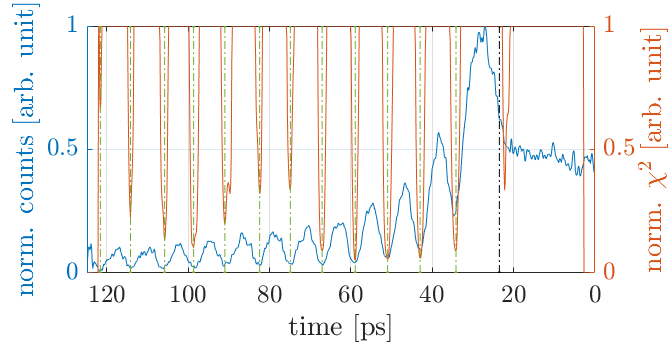}
            \caption{\label{label} Determination of the beginning and end of micro-bunches with the central projection of the modulated bunch (blue solid line), the ionizing laser pulse timing (black dashed line), the weighted distance squared function $\chi^2$ (orange solid line) and its minima (green dashed line)}
            \label{fig:findDefocusedsc} 
        \end{minipage} 
    \end{figure}

For further analysis we first determine the longitudinal (temporal) center of the micro-bunches. The central projection ($-0.1 \, \textnormal{mm} \le x \le 0.1 \, \textnormal{mm}$) of Figure \ref{fig:example_streakcameraimage_laseron} is shown in Figure \ref{fig:findMBsc} and \ref{fig:findDefocusedsc} with the blue solid line. For determining the center of the micro-bunches (Figure \ref{fig:findMBsc}), we fit a second order polynomial $f(t_i,\vec{\lambda}) = \lambda_1 + \lambda_2 \, t_i + \lambda_3 \, {t_i}^2$, with start points $\vec{\lambda} = \{ \lambda_1, \lambda_2, \lambda_3 \}$, constraining $\lambda_3<0$, i.e. a downwards opened parabola to the bunch projection. We let the fit move along the projection centered at time $t_i$ and fit over a range $\{ t_i - \Delta t :t_i+ \Delta t \}$ and $\Delta t =  2.7 \, \textnormal{ps}$, to include most of the data points of a micro-bunch, but avoid covering more than one bunch for the given plasma wakefield period ($T_{pe} = 7.9 \, \textnormal{ps}$ at $n_{pe}= 2\cdot 10^{14} \, \textnormal{cm}^{-3}$). The weighted distance squared function $\chi^2$, giving the difference between the model expectation $f(t_i|\vec{\lambda})$ and the measured projection $y_i$ is given by
    \begin{equation}
        \chi^2 = \sum_i \frac{(y_i - f(t_i|\pmb{\vec{\lambda}}))^2}{\omega_i^2}.
    \end{equation}
We weight the fit with the curvature of the parabola $\omega_i = \lambda_3$, as we expect the strongest curvature in the center of the micro-bunch. The result of the $\chi^2$ fit is shown with the orange solid line in Figure \ref{fig:findMBsc}, restricting the plot to the low values of the function for better visualization. The temporal center of the micro-bunches is defined as the minima of the function, indicated with the green vertical dashed lines.\\
We use a similar analysis but constraining the curvature fit parameter to $\lambda_3>0$, i.e. an upwards opened parabola, to find the minimum between two micro-bunches, corresponding to the maximum defocused regions, see Figure \ref{fig:findDefocusedsc}. Figures \ref{fig:findMBsc} and \ref{fig:findDefocusedsc} show that this automatic procedure finds the micro-bunch center, as well as their beginning and end.\\

\subsection{Micro-Bunch Size Determination}
    \begin{figure}[h!]
    \centering
        \hspace{4pc}%
        \includegraphics[width=16pc,trim={0cm 0cm 0cm 0cm},clip]{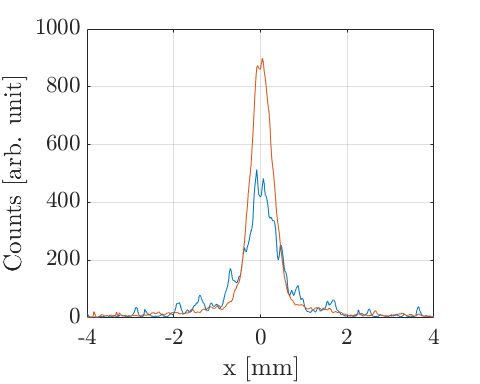}
        \hspace{2pc}%
        \begin{minipage}[b]{15pc}
            \caption{\label{label} Transverse intensity profile of the first micro-bunch in Figure \ref{fig:example_streakcameraimage_laseron} (red line) and of the unmodulated bunch in Figure \ref{fig:example_streakcameraimage_laseroff} (blue line), averaged over $t = 28 \, (\pm \, 0.4) \, \textnormal{ps}$}\
            \label{fig:exampleProfileMB}
        \end{minipage}
    \end{figure}
    
  \begin{figure}[h!]
    \centering
        \includegraphics[width=24pc,trim={1.5cm 0cm 2cm 0cm},clip]{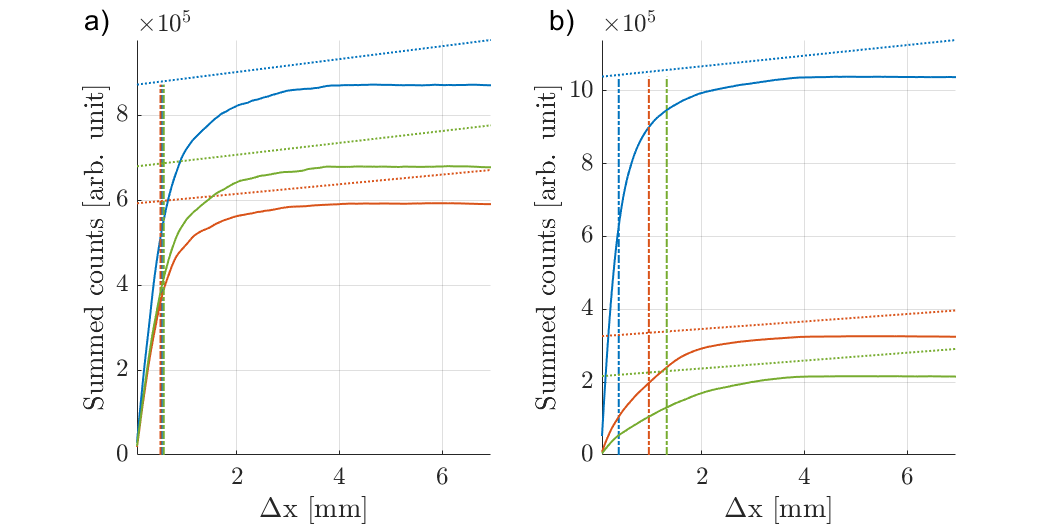}
        \hspace{2pc}%
        \begin{minipage}[b]{11pc}
        \caption{\label{label} Transverse running sum of counts for the unmodulated (a) and modulated (b) bunch over the time range of micro-bunch number one (blue solid line), four (red solid line) and nine (green solid line). The vertical dashed lines show the micro-bunch width, where the sum reaches $60 \%$ of the final value.}
        \label{fig:transverseWidthprofiles}
        \end{minipage}
    \end{figure}
We use the temporal center of the micro-bunches and defocused areas to determine the transverse and longitudinal width of the micro-bunches. A transverse profile of the first micro-bunch of Figure \ref{fig:example_streakcameraimage_laseron} (at $t= 28\, \textnormal{ps}$, obtained from Figure \ref{fig:findMBsc}, averaged over $\pm 0.4 \, \textnormal{ps}$) is shown in Figure \ref{fig:exampleProfileMB} as an example, demonstrating the typical transverse shape of the micro-bunches.
The profiles suggest that there is more charge in the micro-bunch (red line) than in the incoming bunch (blue line) over the same time range. This is not possible, since the proton bunch is strongly relativistic, i.e. protons cannot move longitudinally with respect to each other. This is a good illustration of the slit effect. The micro-bunch width is less that that of the incoming bunch, thus more light is collected through the slit, giving the impression that it contains more charge. Instead, only its charge density is larger (see below).\\
%To avoid assuming a certain shape of the bunch profiles, we determine their size with the help of profiles $W(\Delta x)$ and $L(\Delta t)$, varying the summation range $\Delta x$ and $\Delta t$ in the following.\\
%For the transverse width we calculate $H(\Delta x) = \sum_0^{\Delta x} \sum_{t_{d,j}}^{t_{d,j+1}} I(\Delta x, t)$ (see Figure \ref{fig:transverseWidthprofiles}) with $t_{d,j}$ the temporal position of the $j$th defocused area, determined in Figure \ref{fig:findDefocusedsc}. For the longitudinal (temporal) width we calculate $H(\Delta t) = \sum_{t_{j}}^{t_{d,j}} \sum_{-\Delta x}^{+\Delta x} I(x,\Delta t) $ for the length before the micro-bunch center (Figure \ref{fig:longitudinalWidthprofiles} a, b), and $H(\Delta t) = \sum_{t_{j}}^{t_{d,j+1}} \sum_{-\Delta x}^{+\Delta x} I(x,\Delta t) $ for the length after the center (Figure \ref{fig:longitudinalWidthprofiles} c, d). We sum over $\Delta x = 0.07 \, \textnormal{mm}$ for a less noisy profile. When the micro-bunch is fully covered by the summation range and does not overlap with a second bunch (longitudinally), the summation value saturates $H_{Sat}$ when further increasing the range. We determine the transverse and longitudinal position where $H(\Delta x),H(\Delta t) = 0.6 \cdot H_{Sat}$ as the transverse and longitudinal widths of the micro-bunch.\\
In order to avoid having to assume a transverse (or longitudinal) profile for the micro-bunches, we plot the running sum of counts over each micro-bunch.
For the transverse width we calculate the running sum of counts from the bunch center ($x=0$, Figure \ref{fig:exampleProfileMB}) to the edge of the image over the time range of each micro-bunch, as determined in Figure \ref{fig:findDefocusedsc}. We use the time ranges of the micro-bunches of the modulated bunch to calculate the corresponding sums in the unmodulated bunch. 
%For the transverse width we calculate $W(\Delta x) = \sum_0^{\Delta x} \sum_{t_{d,i}}^{t_{d,i+1}} I(\Delta x, t) \, dt \, dx$, the running sum of counts over each micro-bunch time range as determined on Figure \ref{fig:findDefocusedsc} with $t_{d,i}$ the temporal position of the $i$th defocused area, starting at the bunch axis, as shown on Figure \ref{fig:transverseWidthprofiles}. 
Figure \ref{fig:transverseWidthprofiles}a) shows the profiles of the unmodulated bunch (Figure \ref{fig:example_streakcameraimage_laseroff}) for comparison, and Figure \ref{fig:transverseWidthprofiles}b) of the modulated bunch (Figure \ref{fig:example_streakcameraimage_laseron}). We linearly fit the profiles for $\Delta x>4 \, \textnormal{mm}$, corresponding to summation of background, as the light collection from the proton bunch is limited by the imaging aperture. The subtraction of the linear function (dashed lines in Figure \ref{fig:transverseWidthprofiles}) leads to saturation of the profiles. We define the bunch width (radius) as the radial position at which the sum reaches $60 \%$ of the final value, indicated with the vertical dash-dotted lines. As expected, the width along the unmodulated bunch remains essentially constant, see \ref{fig:transverseWidthprofiles}a). For the modulated bunch, \ref{fig:transverseWidthprofiles}b) shows the running sum over the micro-bunch number one, four and nine, also representing the shape over the other micro-bunches. Unlike the unmodulated bunch, the width (vertical lines) of the individual micro-bunches is changing along the bunch.\\
    \begin{figure}[h!]
    \centering
            \includegraphics[width=16pc]{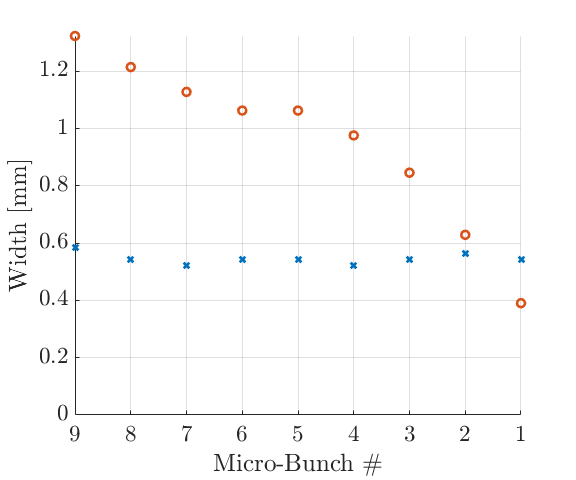}
       \hspace{2pc}%
        \begin{minipage}[b]{11pc}
            \caption{\label{label} Transverse width of the first nine micro-bunches of the modulated bunch (red circles) and the unmodulated bunch (blue crosses), determined with the procedure shown in Figure \ref{fig:transverseWidthprofiles}.}
            \label{fig:AllWidthsMicroBunches}
        \end{minipage}
    \end{figure}
The transverse width of each micro-bunch along the bunch is plotted in Figure \ref{fig:AllWidthsMicroBunches} (red circles) and compared to the width of the incoming bunch (blue crosses) with a mean of $540\,  ( \pm $ \SI{20 }{\micro\m}). One can see that the width of the signal is increasing along the image.\\ %The increase in size of the micro-bunches along the bunch at the OTR foil, after $3.5 \, \textnormal{m}$ propagation in vacuum, could be from stronger focusing along the bunch, thus a smaller size at the plasma exit and/or with a larger divergence, both making the micro-bunches wider at the diagnostic.\\

    \begin{figure}[h!]
    \centering
        \includegraphics[width=24pc,trim={1.5cm 0cm 2cm 0cm},clip]{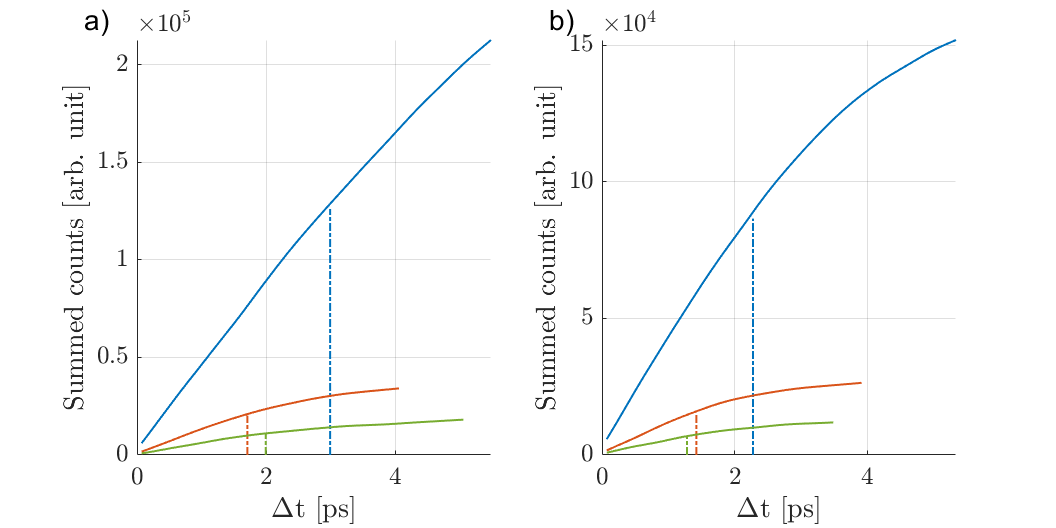}
        \hspace{2pc}%
        \begin{minipage}[b]{11pc}
            %\caption{\label{label} Length determination of the micro-bunches using the summed profiles $L(\Delta t) = \sum_{t_{j}}^{t_{d,j}} \sum_{-\Delta x}^{+\Delta x} I(x,\Delta t)$ before (a, b) and $L(\Delta t) = \sum_{t_{j}}^{t_{d,j+1}} \sum_{-\Delta x}^{+\Delta x} I(x,\Delta t) $ after (c, d) the micro bunch center $t_j$, with $\Delta x = 0.07 \, \textnormal{mm}$. The different colors correspond to the individual micro bunches, the vertical lines show the temporal widths $l_{t,b}$ before and $l_{t,a}$ after the micro-bunch, where $L(\Delta t) = 0.6 \cdot L_{Sat}$.}
            \caption{\label{label} Longitudinal running sum of counts from each micro-bunch center to its beginning (a) or end (b) for micro-bunch number one (blue solid line), four (red solid line) and nine (green solid line). The vertical dashed lines show the micro-bunch length, where the sum reaches $60 \%$ of the final value.}
            \label{fig:longitudinalWidthprofiles}
        \end{minipage}
    \end{figure}    
We use a similar approach to determine the length of the micro-bunches, see Figure \ref{fig:longitudinalWidthprofiles}. The micro-bunch length before and after the micro-bunch center here differ from each other and are thus treated individually. 
%We calculate $L(\Delta t) = \sum_{t_{i}}^{t_{d,i}} \sum_{-\Delta x}^{+\Delta x} I(x,\Delta t)  \, dx \, dt$ for the length of the $i$th micro-bunch from the center $t_i$ to the previous defocused area $t_{d,i}$ (Figure \ref{fig:longitudinalWidthprofiles} a), and $L(\Delta t) = \sum_{t_{i}}^{t_{d,i+1}} \sum_{-\Delta x}^{+\Delta x} I(x,\Delta t)  \, dx \, dt$ for the length to the subsequent defocused area $t_{d,i+1}$ (Figure \ref{fig:longitudinalWidthprofiles} b). 
We calculate the running sum of counts from the center of the micro-bunch (Figure \ref{fig:findMBsc}) to its beginning and end (Figure \ref{fig:findDefocusedsc}). For the given measurement, the bunch is not fully modulated near the ionization front, thus the counts between the micro-bunches do not reach the value 0, i.e. the sums do not saturate. Therefore we limit the range of summation with the beginning (\ref{fig:longitudinalWidthprofiles}a) and end (\ref{fig:longitudinalWidthprofiles}b) of the micro-bunch and the time of the ionizing laser pulse as the beginning of the first micro-bunch. Here we sum over the transverse range $\SI{-70}{\micro\m} <x< \SI{70}{\micro\m} $ for a less noisy profile. The vertical dash-dotted lines indicate the determined micro-bunch length, where the sum reaches $60 \%$ of the final value.\\
%of the $i$th micro-bunch, $l_{i,b}$ before and $l_{i,a}$ after the micro-bunch center, where the profiles reach $L(\Delta t) = 0.6 \cdot L_{Sat}$.\\
    \begin{figure}[h!]
    \centering
            \includegraphics[width=16pc,trim={0.5cm 0cm 0cm 0cm},clip]{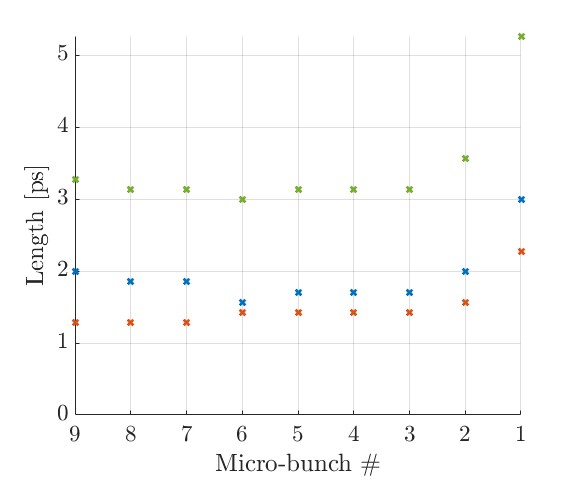}
        \hspace{2pc}%
        \begin{minipage}[b]{11pc}
            \caption{\label{label} Length of the first nine micro-bunches from the micro-bunch center to its beginning (blue crosses) or end (red crosses), determined with the procedure shown in Figure \ref{fig:longitudinalWidthprofiles}, and their sum (green crosses)}
            \label{fig:AllLengthsMicrobunches}
        \end{minipage}
    \end{figure}
The lengths are summarized in Figure \ref{fig:AllLengthsMicrobunches} for each micro-bunch along the bunch. The length from the center to the beginning as blue, from the center to the end as red, and the sum of the two as green crosses. One can see that the first two micro-bunches are longer and after the third micro-bunch the length saturates to $3.1 \, (\pm \, 0.1) \, \textnormal{ps}$. In the following we use the length of the micro-bunches for the unmodulated bunch for the comparison of charge in a given length along the bunch. Note that temporal resolution might lower the counts per pixel for a signal with time structure, as the modulated bunch, while it should not affect a signal without, as the unmodulated bunch.\\

\section{Results}
\label{sec:results} 
Since the proton bunch is cylindrically symmetric, its image onto the streak camera slit is also symmetric. With the minimum bunch diameter (\SI{780}{\micro\m} at the OTR screen from Figure \ref{fig:AllWidthsMicroBunches}, corresponding to \SI{220}{\micro\m} at the streak camera slit due to the de-magnification by the OTR light transport) being larger than the slit width (\SI{20}{\micro\m}), the streak camera image profile at each time (Figure \ref{fig:exampleProfileMB}) can be interpreted as a measurement of the bunch charge density as a function of time $n(r,t)$ or $n(x,t)$ on the images. The charge at each time of the image, or in each micro-bunch as determined above, can be calculated multiplying the charge density by $2 \pi \, r \, dr$. 

\subsection{Charge Determination of the Proton Bunch}
    \begin{figure}[h!]
        \begin{minipage}{18pc}
            \includegraphics[width=16pc]{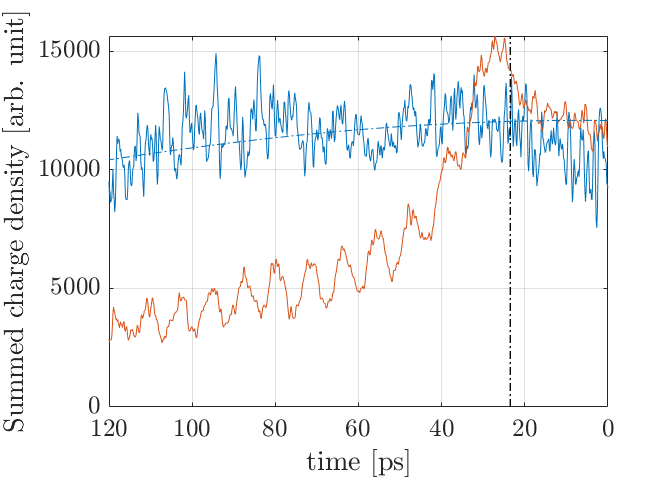}
            \caption{\label{label} Charge density summed transversely over the unmodulated bunch (blue solid line) and over the modulated bunch (red solid line), theoretical Gaussian profile (blue dashed line) and timing of ionizing laser pulse (black dashed line)}
            \label{fig:TotalChargeNormOff}
        \end{minipage}\hspace{2pc}%
        \begin{minipage}{18pc}
            \includegraphics[width=16pc,trim={0cm 0cm 0cm 0cm},clip]{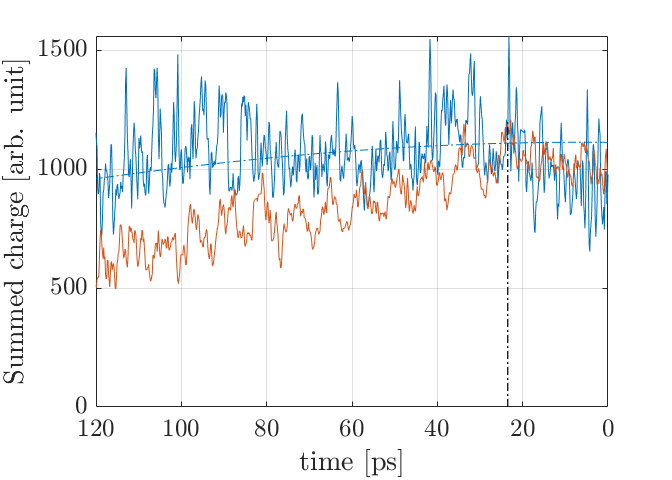}
            \caption{\label{label} Charge summed transversely over the unmodulated bunch (blue solid line) and over the modulated bunch (red solid line), theoretical Gaussian profile (blue dashed line) and timing of ionizing laser pulse (black dashed line)}
            \label{fig:TotalChargeNormOn}
        \end{minipage} 
    \end{figure}
We apply this procedure to calculate the charge of the image of the unmodulated (Figure \ref{fig:example_streakcameraimage_laseroff}) and modulated bunch (Figure \ref{fig:example_streakcameraimage_laseron}). To avoid the signal from the timing reference laser pulse we use only half of the image ($x > 0$).
We compare the charge density summed transversely over the entire modulated bunch (red curve) with the unmodulated bunch (blue curve) in Figure \ref{fig:TotalChargeNormOff}. As expected, the shape for the unmodulated bunch follows the Gaussian bunch distribution (with length $\sigma_{\zeta} = 300 \, \textnormal{ps}$ and $t=0$ being $6 \, \textnormal{ps}$ behind the bunch center and amplitude normalized to the measured profile), indicated with the blue dashed line. The summation of the modulated bunch includes focused and defocused protons. 
%The resulting curves Figure \ref{fig:TotalChargeNormOff} show that the charge density along the bunch is decreasing significantly, when the bunch is modulated (red) in comparison with the unmodulated bunch (blue).\\
Summing the charge density transversely over the bunch before the start of the plasma ($t<24 \, \textnormal{ps}$) leads to similar values for the modulated bunch and the incoming bunch, as expected. Summing the charge density transversely over the bunch within the plasma ($t>24 \, \textnormal{ps}$) shows significantly lower counts of the modulated bunch than the incoming bunch. The sum of the modulated bunch also exhibits the periodic modulation from the self-modulation process.\\
The decrease in signal along the image in Figure \ref{fig:TotalChargeNormOff} for the modulated bunch (red curve) is caused by the increase in width (see Figure \ref{fig:AllWidthsMicroBunches}, red circles). We can account for the effect of the slit and determine the charge of the image by multiplying the images containing the charge density with $2\pi \, r \, dr$. Figure \ref{fig:TotalChargeNormOn} shows the sum of the charge over the modulated bunch (red curve) compared to the incoming bunch (blue curve). 
%As expected from Figure \ref{fig:example_streakcameraimage_laseron}, Figure \ref{fig:TotalChargeNormOff} shows that the charge density of the modulated bunch (red) exhibits the periodic modulation from the self-modulation process. The profile of the modulated bunch shows a higher density for the first plasma period ($24\, \textnormal{ps} \le t \le 32 \, \textnormal{ps}$) than the unmodulated bunch (blue) resulting from transverse focusing of protons within the plasma (see also width of the first micro-bunch in Figure \ref{fig:AllWidthsMicroBunches}). For times $t>34 \, \textnormal{ps}$ the charge density of the modulated bunch is decreasing significantly along the bunch with respect to the unmodulated bunch.\\
%In order to determine the charge from the measured charge density, the width of the light signal, thus fraction of light captured by the slit of the streak camera, needs to be considered. Assuming a cylindrical symmetric signal, and a signal large with respect to the slit width ($20 \, \mu \textnormal{m}$), we can account for the effect of the slit by multiplying the streak camera image with the radial position: The charge $Q$ out of the charge density $I$ in the images is given by $Q = I \, 2 \, \pi \, r \, dr$, with $r \equiv x$ due to the radial symmetry. 
%Figure \ref{fig:TotalChargeNormOn} shows that when computing charge $Q(t) = 2\, \pi \, \sum_r I(r,t)  \, r\, dr$, 
It shows that the charge along the self-modulated bunch is very close to that of the unmodulated bunch, following the Gaussian profile. The procedure recovers the same charge for parts of the bunch before the plasma ($t<24 \, \textnormal{ps}$). However, it retains some of the modulation in the charge density and the recovered charge decreases along the bunch when compared to the incoming bunch charge. These deviations are probably due to light collection and detection limitations of the diagnostic. Protons are more and more defocused along the bunch (see \cite{Marlene}) and images show that they fall out of the imaged field ($-4 \, \textnormal{mm} < x <4 \, \textnormal{mm}$) later along the bunch. Also, the bunch charge density decreases further along the bunch. The streak camera has a limited signal to noise ratio and low level light is not detected, falling below the detection threshold. The effects increase along the bunch.\\ 
%This leads to the images containing the charge of the unmodulated (top) and modulated (middle) bunch in Figure \ref{fig:Charge_norm_On}. The self-modulation of the relativistic bunch is purely transverse \cite{AWAKE}. Thus, if we sum radially the charge $Q$ of the entire image $0<x<4 \, \textnormal{mm}$, covering focused and defocused protons, we expect a similar charge value for the modulated and unmodulated bunch. The transversely summed charge over the entire unmodulated (blue) and modulated (red) bunch is shown in Figure \ref{fig:TotalChargeNormOn}, showing similar values, as expected.\\
%After several modulation periods (here at $t \gtrsim 80 \, \textnormal{ps}$) the summed charge of the modulated bunch deviates from the unmodulated bunch, as the defocusing of protons is increasing and thus the emitted light is not be captured by the OTR transport optics. Additionally, the difference between the values of the curves as well as the remaining modulation of the curve for the modulated bunch, can be explained considering the detection threshold of the streak camera. When the protons are defocused, the light signal covers a larger radial range, thus potentially with a larger extend of a signal below the camera threshold. 
Figure \ref{fig:TotalChargeNormOn} shows how well the charge along the bunch can be determined with this diagnostic and procedure. Now, that we have developed a procedure, to recover the charge in the bunch from the measurement of its charge density, and determined its limitation, we can measure the charge carried by each micro-bunch, i.e. not including the charge of defocused protons, and compare it to the incoming charge.\\

\subsection{Charge Determination of Individual Micro-Bunches}
    \begin{figure}[h!]
        \begin{minipage}{18pc}
            \includegraphics[width=18pc,trim={0cm 0.5cm 0cm 0cm},clip]{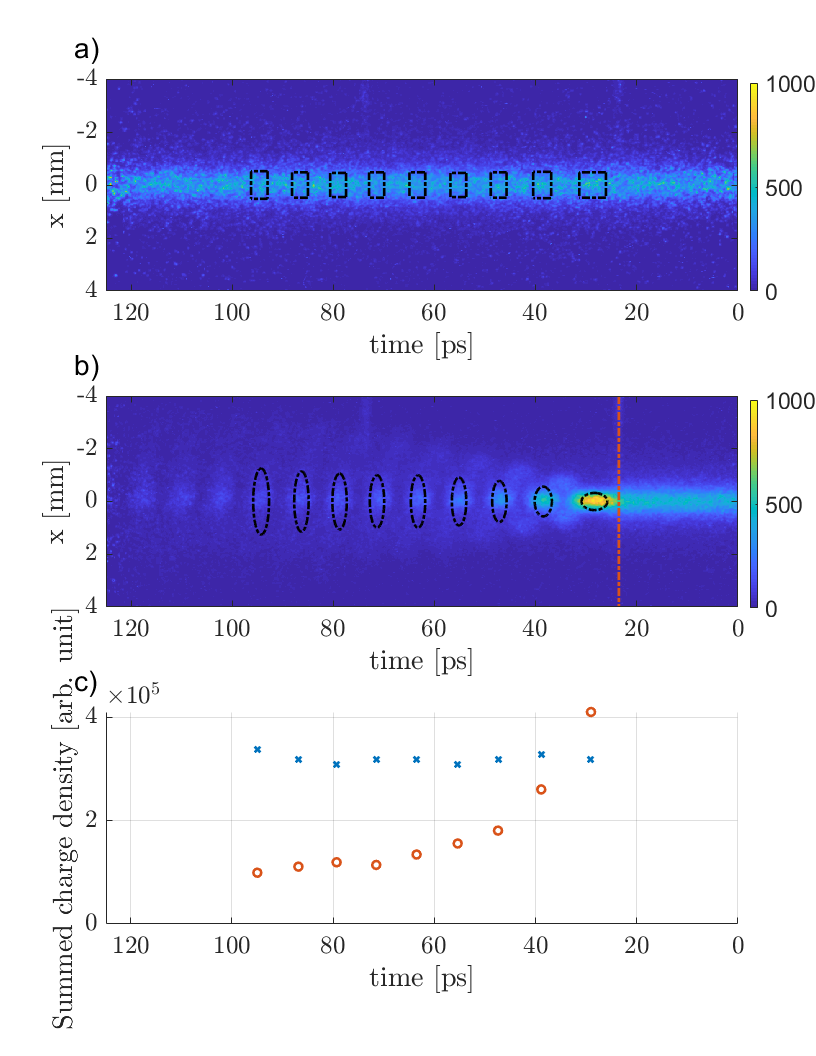}
            %\caption{\label{label} The size of micro-bunches are shown in a) for the unmodulated bunch and b) for the modulated bunch on the original streak camera images representing the charge density. The charge density summed over the micro-bunches $ \frac{1}{(w_{t,b}+w_{t,a})} \sum_{-w_r}^{w_r} \sum_{-w_{t,b}}^{w_{t,a}} \, I(r,t) $ is given in c) for the unmodulated (blue crossses) and the modulated (red circles) bunch, compared with the Gaussian profile (blue dashed line)}
            \caption{\label{label} Streak camera images of the unmodulated (a) and modulated (b) bunch. The squares in a) represent the width of the unmodulated bunch and the length of the micro-bunches, the ellipses in b) show the widths and length of the micro-bunches. Summing the counts of the image, representing the charge density $n(r,t)$, over the indicated squares or ellipses leads to c) for the unmodulated (blue crosses) and the modulated (red circles) bunch.}
            \label{fig:Charge_norm_Off}
        \end{minipage}\hspace{2pc}%
        \begin{minipage}{18pc}
            \vspace{-0.4cm}
            \includegraphics[width=18pc,trim={0cm 0.5cm 0cm 0cm},clip]{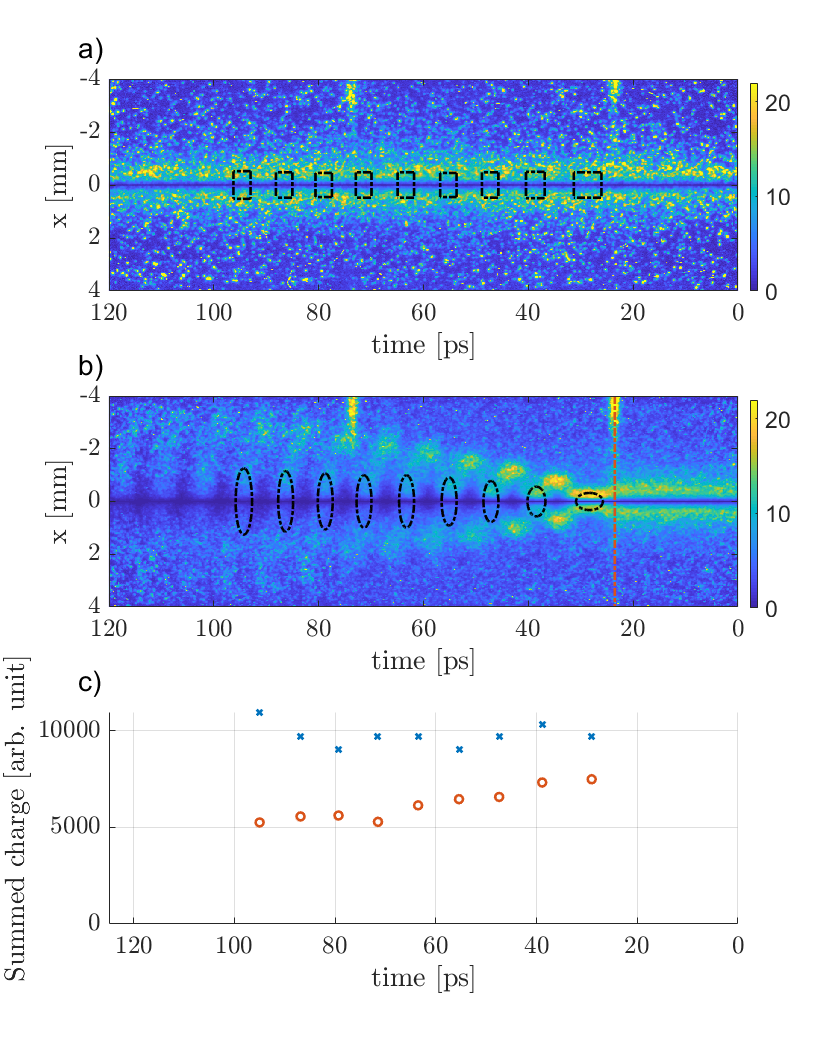}
            %\caption{\label{label}  The size of micro-bunches are shown in a) for the unmodulated bunch and b) for the modulated bunch on the processed streak camera images representing the charge. The charge summed over the micro-bunches $\frac{Q}{(w_{t,b}+w_{t,a})}=\frac{ 2\pi }{(w_{t,b}+w_{t,a})} \sum_{-w_{r}}^{w_{r}} \sum_{-w_{t,b}}^{w_{t,a}} \, I(r,t) \, r \,dr$ is given in c) for the unmodulated (blue crossses) and the modulated (red circles) bunch, compared with the Gaussian profile (blue dashed line)}
            \caption{\label{label} The streak camera images multiplied by $2 \pi \, r \, dr$ revealing the charge is shown in a) for the unmodulated and in b) for the modulated bunch. The squares in a) indicate the width of the unmodulated bunch and length of the micro-bunches, the ellipses in b) the width and length of the micro-bunches. Figure c) shows the charge $n(r,t) \, 2 \pi \, r \, dr$ summed over the squares (blue crosses) and ellipses (red circles).}
            \label{fig:Charge_norm_On}
        \end{minipage} 
    \end{figure}
We use the transverse width and length of the incoming bunch and the micro-bunches, as obtained using the procedure detailed in section \ref{sec:method} (Figures \ref{fig:AllWidthsMicroBunches} and \ref{fig:AllLengthsMicrobunches}), to determine the relative charge per micro-bunch. For comparison of the charge in each micro-bunch with the charge in the incoming bunch, we use the same time ranges for the summation on the images of the modulated and unmodulated bunch.\\
Summation of the charge density, as given by the original streak camera image, over the range of the micro-bunches, is shown in Figure \ref{fig:Charge_norm_Off}. %The sum for micro-bunch $i$ normalized to its temporal length is calculated as $\frac{1}{(l_{i,b}+l_{i,a})} \sum_{-w_i}^{w_i} \sum_{-l_{i,b}}^{l_{i,a}} \, I(r,t) \, dt \, dr$.
The image of the unmodulated bunch, where the squares indicate the transverse width of the unmodulated bunch and the length of the micro-bunches, is shown in a); the image of the modulated bunch, with the ellipses of the width and length of the micro-bunches in b). In c) the charge density summed over the squared areas in a) (blue crosses) and over the ellipses in b) (red circles) is shown. One can see that the summed charge density for the unmodulated bunch remains essentially constant (considering the limitation of this procedure, the longitudinal bunch position close to the center, and the difference in length for the first micro-bunches and thus summation length being small). In constrast, the summed charge density of the micro-bunches decreases rapidly along the bunch, due to the increasing radial size of the signal (see Figure \ref{fig:AllWidthsMicroBunches}). This is consistent with Figure \ref{fig:TotalChargeNormOff}, where the charge density of the unmodulated bunch follows the Gaussian bunch distribution, while the charge density of the modulated bunch decreases along the bunch.\\
%We can compare the values with the expected longitudinal profile of the proton bunch (blue dotted line), assuming a Gaussian profile, the position along the bunch (here: ionizing laser pulse at $t = 24 \, \textnormal{ps}$ in figure, being $30 \, {ps}$ ahead of the center of the proton bunch with $\sigma_{\zeta} = 300 \, \textnormal{ps}$) and the height of the peak normalized to the mean value of the blue crosses. 
%The errorbars of the summed counts per fictional micro-bunch for the unmodulated bunch are obtained by the standard deviation of the difference to the blue dotted line. 
%This can be compared to the summed counts of the micro-bunches of the modulated bunch. While the summed charge density for the unmodulated bunch follows the Gaussian profile, one can see that the summed charge density of the micro-bunches is decreasing rapidly along the bunch, due to the increasing radial size of the signal (see Figure \ref{fig:AllWidthsMicroBunches}). This is consistent with Figure \ref{fig:TotalChargeNormOff}, where the charge density of the unmodulated bunch follows a Gaussian bunch distribution, while the charge density of the modulated bunch is decreasing along the bunch.\\
In order to determine the charge per micro-bunch we multiply the streak camera images $n(r,t)$ by $2\pi \, r \, dr$, as shown in Figure \ref{fig:Charge_norm_On}a) for the unmodulated and b) for the modulated bunch. We sum over the same squares and ellipses as described above, in order to determine the charge per micro-bunch.
%The charge per micro-bunch $i$ normalized to its temporal length is calculated as $\frac{Q}{(l_{i,b}+l_{i,a})}=\frac{ 2\pi }{(l_{i,b}+l_{i,a})} \sum_{-w_i}^{w_i} \sum_{-l_{i,b}}^{l_{i,a}} \, I(r,t) \, r \, dt \,dr$. 
Figure c) compares the charge per micro-bunch with the charge of the incoming bunch. Again, we expect the charge of the unmodulated bunch to be essentially constant (central position within the long Gaussian bunch and small changes in micro-bunch length), which is confirmed by the measurement in blue. In red it is demonstrated that also the charge per micro-bunch is roughly constant along the bunch. The mean charge per micro-bunch covered in the ellipse is $64 \, ( \pm \, 9 ) \%$ of the charge covered in the squares of the incoming bunch. 

\section{Summary}
\label{sec:conclusion}
We showed that because of the streak camera slit the streak camera images must be interpreted as charge density of the proton bunch after $3.5 \, \textnormal{m}$ of propagation in vacuum and not in the plasma. We presented a procedure that recovers the charge in the bunch from the streak camera images. Applying the procedure we demonstrated that the charge in each micro-bunch is constant along the bunch for the first nine micro-bunches. The charge in the micro-bunches corresponds to more than $60 \%$ of the charge of the incoming bunch, over the same time period, within limitations of the diagnostic. This procedure will be used to characterize the result of the self-modulation process on the proton bunch and potentially for studying the resulting wakefields.

%We demonstrate that the charge per micro-bunch along the self-modulated proton bunch after $10 \, \textnormal{m}$ of propagation in plasma is constant. After the $2.5 \, \textnormal{m}$ of propagation in vacuum to the OTR foil, the radii of the micro-bunches are increasing along the modulated bunch, thus the charge density inside the micro-bunches is decreasing. Further studies are necessary to explain whether the expansion is occurring during the propagation inside plasma, or whether the divergence of the micro-bunches is increasing along the modulated bunch, leading to a larger radius after a certain propagation distance in vacuum after the plasma exit.\\
%As the proton bunch is the driver of the wakefield, used for the acceleration of electrons, we aim for a maximum loss of proton energy. Thus, from linear theory, ideally the micro-bunch length would correspond to $1/4$th of the plasma period, placed in the focusing and decelerating phase, avoiding protons in the accelerating phase gaining energy. For high wakefield amplitudes a high charge within this period, thus no decrease of charge per micro-bunch along the entire bunch, is favorable. 

\section{Acknowledgments}
\label{sec:Acknowledgments}
    This work is sponsored by the Wolfgang Gentner Program of the German Federal Ministry of Education and Research (05E15CHA).

\end{document}